\documentclass[10pt, doublecolumn]{IEEEtran}
\usepackage{graphicx}
\usepackage{caption}
\usepackage[ruled]{algorithm2e}
 
\ifCLASSOPTIONcompsoc
  \usepackage[caption=false,font=normalsize,labelfont=sf,textfont=sf]{subfig}
\else
  \usepackage[caption=false,font=footnotesize]{subfig}
\fi
\usepackage{indentfirst}
\usepackage{graphicx}              
\usepackage{subcaption}            
\usepackage{graphicx}
\usepackage{bm}
\usepackage{amsmath}
\allowdisplaybreaks[4]
\usepackage{amssymb}
\usepackage{times}
\usepackage{mathtools}
\usepackage{psfrag}
\usepackage{cite}
\usepackage{lastpage}
\usepackage{fancyhdr}
\usepackage{color}
\usepackage{amsthm}
\usepackage{bigints}

\theoremstyle{remark}

\begin{document}
\title{Pinching Antennas: Principles, Applications and Challenges}
\author{Zheng Yang,~\IEEEmembership{Member, IEEE}, Ning Wang, Yanshi Sun,~\IEEEmembership{Member, IEEE},  Zhiguo Ding,~\IEEEmembership{Fellow, IEEE},  Robert Schober,~\IEEEmembership{Fellow, IEEE},  George K. Karagiannidis,~\IEEEmembership{Fellow, IEEE},      Vincent W. S. Wong,~\IEEEmembership{Fellow, IEEE}, and 
 Octavia A. Dobre,~\IEEEmembership{Fellow, IEEE}

\thanks{Z. Yang is with the Fujian Provincial Engineering
Technology Research Center of Photoelectric Sensing Application, Fujian Normal University, Fuzhou 350117, China. N. Wang is with the School of Information and Control Engineering, China University of Mining and Technology, Xuzhou, 221116, China. Y. Sun is with the School of Computer Science and Information Engineering, Hefei University of Technology, Hefei, 230009, China. Z. Ding is with Khalifa University, Abu Dhabi, UAE, and the University of Manchester, Manchester, M1 9BB, UK. R. Schober is with the Institute for
Digital Communications, Friedrich-Alexander-University Erlangen-Nurnberg (FAU), Germany. 
G. K. Karagiannidis is with the Wireless Communications and Information Processing (WCIP) Group, Electrical \& Computer Engineering Dept., Aristotle University of Thessaloniki, 54 124, Thessaloniki, Greece. V. W. S. Wong is the with Department of Electrical and Computer Engineering, The University of British Columbia, Vancouver,  Canada. O. A. Dobre is with the Department of Electrical and Computer Engineering, Memorial University, St. John’s, NL, Canada.  }
}
\maketitle
\begin{abstract}
Flexible-antenna systems, such as  fluid antennas and movable antennas, have been recognized as key enabling technologies for sixth-generation (6G) wireless networks, as  they can intelligently reconfigure the effective channel gains of the users and hence  significantly improve their data transmission capabilities. However, existing flexible-antenna systems have been designed to combat small-scale fading in non-line-of-sight (NLoS) conditions. As a result, they lack the ability to establish  line-of-sight links, which are typically 100 times stronger than NLoS links. In addition, existing flexible-antenna systems have limited flexibility, where adding/removing an antenna is not straightforward.
This article introduces an innovative flexible-antenna system called pinching antennas, which are realized by applying small dielectric particles to waveguides.  We first describe the basics of pinching-antenna systems and their ability to provide strong LoS links by deploying pinching antennas close to the users as well as their capability to scale  up/down the antenna system. We then focus on communication scenarios with different numbers of waveguides and pinching antennas, where innovative approaches to implement multiple-input multiple-output and non-orthogonal multiple access are discussed. 
In addition, promising 6G-related  applications of pinching antennas, including integrated sensing and communication and next-generation multiple access, are presented. Finally, important directions for future research, such as waveguide deployment and channel estimation, are highlighted.

\end{abstract}

\section{Introduction}

The sixth-generation (6G) communication networks are  designed to support ultra-high-speed data transmission, provide seamless connectivity, and enable massive device density in diverse wireless environments\cite{Zhang19vtm6g}. The achievable data rate of a communication link depends primarily on the channel conditions between the transceivers, which are affected by both small-scale and large-scale fading.  Multiple-input multiple-output (MIMO) systems can mitigate the adverse effects of small-scale fading by  exploiting the multi-path propagation and spatial diversity inherent to wireless channels \cite{Zhang19vtm6g}. Recently, several advanced flexible antenna systems, such as fluid antennas, and movable antennas, have been developed to dynamically reconfigure the effective channel gains caused by multi-path fading between  transmitter and receiver. 

 Fluid antennas use reconfigurable materials, such as liquid metals or ionized solutions, and their positions can be modified in real-time, providing enhanced signal transmission flexibility for space-limited devices \cite{Wong21TWCFAS}. Movable antennas are capable of dynamically adjusting their  position to alter the direction and coverage of their radiation, allowing them to effectively adapt to variations of  user locations  \cite{Zhutwc24MA}. Both  fluid antennas and  movable antennas  limit antenna movement to a few wavelengths, which has an insignificant impact on the large-scale path loss. For example, in cases where the LoS link is blocked, adjusting the transceiver antenna positions by moving them over a few wavelengths is often insufficient to restore the LoS link, a limitation  particularly pronounced at high carrier frequencies and associated shorter wavelengths. 
We note that the performance of a communication system is expected to be degraded if there is no LoS link, since the strength of a LoS link is generally much larger than that of a non-line-of-sight (NLoS) link. In addition, there is limited flexibility for the configuration of existing flexible-antenna systems, where adding or removing antenna elements is neither straightforward nor cost-effective.

Pinching antennas, first introduced by DOCOMO in 2022 \cite{suzuki2022pinching}, are an innovative solution that combines low-cost deployment with high adaptability. Conceptually, pinching antennas are realized by using small dielectric particles to activate specific points along a waveguide. Thus, a pinching antenna acts as a leaky wave antenna, creating controlled radiation sites \cite{Zhiguo24pinchinga}, \cite{Chongjun24pinchinga}. Unlike traditional fixed-location antennas, pinching antennas allow flexible positioning without additional hardware, making them feasible and practical to deploy in desired locations to establish adjustable and reliable LoS transceiver links. This approach provides precise control over complex communication environments, such as supporting links in dynamic or obstructed locations. 

Motivated by the aforementioned discussion, this article aims to introduce the principles of pinching antennas for different waveguide deployments, examine potential system designs, and highlight promising directions for future research. Section II introduces the basics of radio propagation, waveguides, and pinching antennas. Section III focuses on communication scenarios using pinching-antenna systems with a single waveguide. Section IV covers communication scenarios using multiple waveguides in pinching-antenna systems. Important 6G-related application scenarios of pinching antennas are elaborated in Section V, followed by promising future research directions in Section VI.  Finally, Section VII concludes the paper.

\begin{figure*}[!t]
\vspace{0em}
\setlength{\abovecaptionskip}{0em}   
\setlength{\belowcaptionskip}{0em}   
\centering
\includegraphics[width=0.9\linewidth]{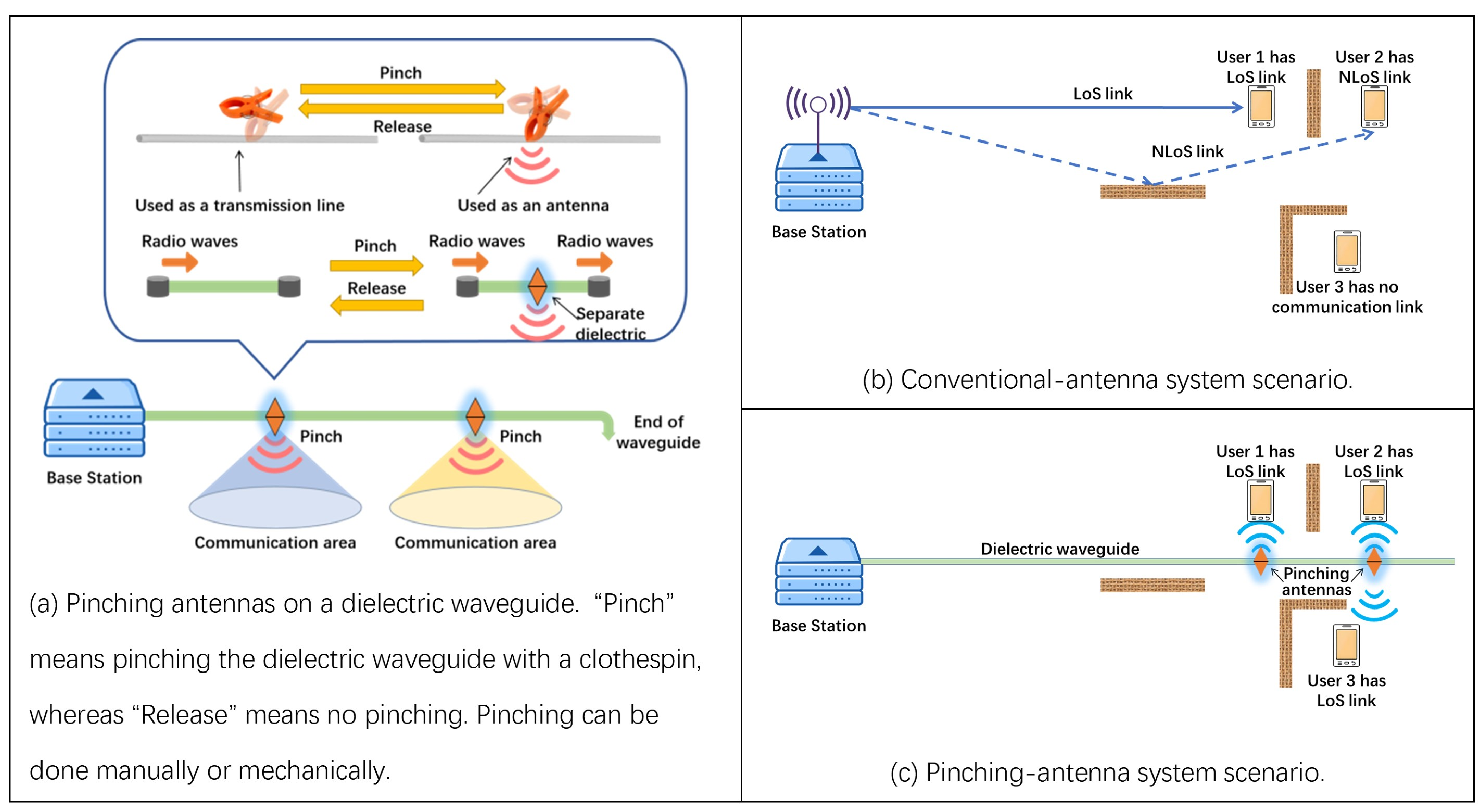}
\vspace*{0.5em}
\caption{(a) Pinching antennas on a dielectric waveguide, where the location of the pinching antennas can be adjusted to create communication zones in the surrounding area; (b) and (c) show the difference between conventional-antenna systems and pinching-antenna systems.}
\label{fig_CommScenario}
\end{figure*}

\section{Radio Wave Propagation and Waveguides}

In this section, we first discuss the characteristics of radio channels and waveguides. Then, we provide an overview of the basic concepts and features of pinching antennas.

\subsection{Radio Channels}
LoS and NLoS are two important types of communication paths that can affect the quality and reliability of wireless transmission. LoS refers to a direct and unobstructed path between transceivers. LoS links have low signal attenuation and distortion, resulting in reliable signal transmission. The probability of having an LoS link is a function of the transmission distance \cite{Zhiguo24pinchinga}, i.e., $\mathbb{P}(\text{LoS}) = e^{-(\rho_{\text{LoS}}} \cdot r)$, where $\rho_{\text{LoS}}$ is a system parameter related to building density and $r$ is the transmission distance. In contrast, NLoS links are caused by obstacles or obstructions, such as buildings, trees, or terrain, that reflect or diffract the signals between the transceivers, resulting in signal degradation and multi-path fading.
In \cite{7434656}, the  channel for  outdoor microcellular scenarios at 28 GHz was modeled and simulated, and it was shown that the path loss of NLoS links is more than 20 dB higher than that of LoS links, and this difference increases with the transceiver distance.

\subsection{Principles of Waveguides}

A waveguide is a specific tubular, strip, or wire-like structure that facilitates the propagation of electromagnetic waves in one direction while restricting their propagation in other directions. For example, a dielectric waveguide is a rod-shaped dielectric material surrounded by another dielectric material. Since the permittivity of the inner dielectric material is much higher than that of the surrounding waveguide material, high-frequency electromagnetic waves are mainly guided within the inner dielectric material. 

When the wave propagates through a waveguide, its wavelength changes accordingly due to the structure and material properties of the waveguide. In dielectric waveguides, the wavelength is shorter than the wavelength  in free-space. This is because the wave in the waveguide is confined by the structure of the waveguide and cannot propagate as freely as in free space. The wavelength of a dielectric waveguide can be calculated as $\lambda_g = \lambda_0/\sqrt{\varepsilon_r}$, where $\lambda_0$ is the wavelength in free space and $\varepsilon_r$ is the relative permittivity of the inner dielectric material, which represents the permittivity of materials relative to the permittivity of free space.

\subsection{Basics of Pinching Antennas}

The main idea of pinching antennas is to use a dielectric waveguide as an antenna by pinching it with a different dielectric material. {For example, dielectric waveguides  may be based on PolyTetraFluoroEthylene (PTFE) with a relative permittivity of 2.1 for the inner conductor and air with a relative permittivity of nearly 1.00 for the outer conductor. Using this waveguide the concept of pinching antennas was experimentally verified in \cite{suzuki2022pinching} by transmitting radio-frequency signals in the 60 GHz band.} In fact, pinching can create radiating radio waves at any point along the dielectric waveguide to establish communication zones in the surrounding area, as shown in Fig. \ref{fig_CommScenario} (a).  The unique feature of pinching-antenna systems is that the location of the pinching antenna can be flexibly adjusted. This feature of pinching-antenna systems enables precise and flexible establishment of communication areas. Compared to conventional fixed-location-antenna systems, pinching antennas can be easily deployed at desirable locations on dielectric waveguides to create  new LoS links or improve existing LoS links. The benefits of pinching-antenna systems can be illustrated by considering different communication scenarios as follows:
\begin{itemize}
\item Even if there is a LoS link in the conventional-antenna system (e.g., user 1 in Fig. \ref{fig_CommScenario} (b)), a pinching antenna can be placed closer to the users, as shown in Fig. \ref{fig_CommScenario} (c). In particular, by placing the pinching antenna next to user 1, the distance between the transmitter and the user can be significantly reduced, ensuring a stronger LoS link. Therefore, the use of pinching antennas is an attractive and practical solution for mitigating large-scale path loss.

\item When there is no LoS link in the conventional-antenna system (e.g., user 2 in Fig. \ref{fig_CommScenario} (b)), the communication between the transceivers is negatively affected by the heavily attenuated NLoS transmission path. By properly adjusting the pinching antenna position, it is possible to create an LoS link between the base station and user 2, as shown in Fig. \ref{fig_CommScenario} (c). In this scenario, the performance gain of pinching-antenna system over the conventional one is expected to be significant, since the path loss of NLoS links is much higher than that of LoS links.

\item In extreme environments, such as areas with tall buildings, enclosed spaces, tunnels, or other significant obstructions, the communication links can be completely blocked. In this scenario, even with high transmit power and many antennas, a conventional communication system cannot provide reliable signal transmission (e.g., user 3 in Fig. \ref{fig_CommScenario} (b)). By adjusting the position of the pinching antenna to a suitable location for user 3, either an LoS or an NLoS link can be established, as shown in Fig. \ref{fig_CommScenario} (c). 
\end{itemize}


\begin{figure*}[!t]
\centering
\includegraphics[width=0.9\linewidth ]{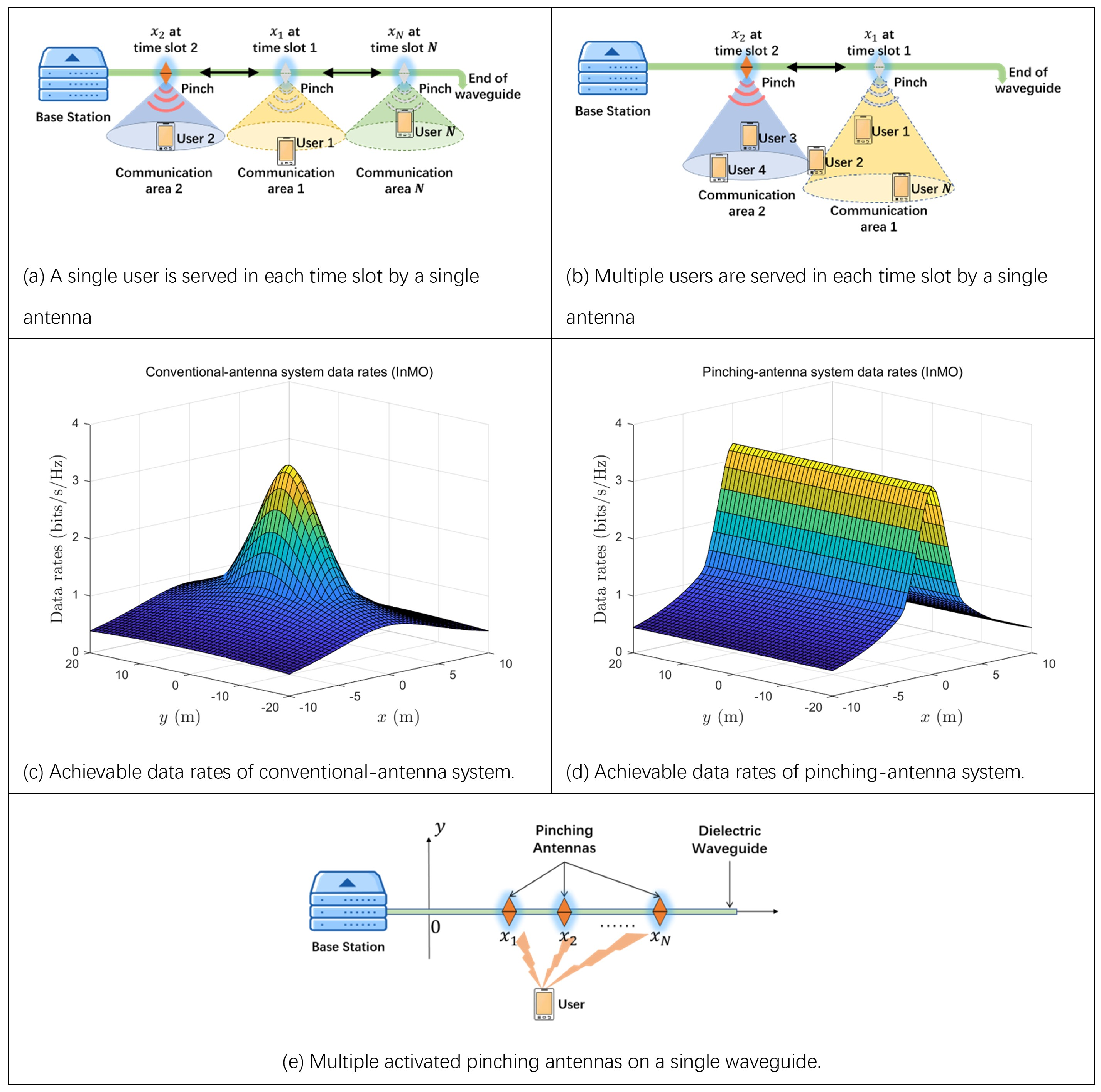}
\caption{Communication scenarios with a single waveguide.   The waveguide of the pinching-antenna systems is deployed along the $y$ axis at a height of  $3$~m, whereas the conventional-antenna system is placed at $(0,0,3)$. The indoor-mixed office (InMO) model \cite{7434656} is used to determine the LoS probability.}
\label{fig_SWSA}
\end{figure*}
 
\section{Communications with a Single Waveguide}

In this section, we focus on the deployment of pinching antennas on a single waveguide, where the application of a single and multiple activated pinching antennas is discussed in the following subsections, respectively.

\subsection{Single Waveguide with a Single Activated Pinching Antenna}

{ When a single pinching antenna is activated on a single waveguide in a multi-user communication scenario, two different multiple access schemes can be used. First, an orthogonal multiple access (OMA) scheme, such as time division multiple access (TDMA), can be used to serve  multiple users.  As shown in Fig. \ref{fig_SWSA}(a), the pinching antenna is activated in $N$ time  slots at $N$ different locations along the waveguide to serve $N$ different users. To simplify the implementation of this scheme, multiple separate dielectric pinches can be pre-deployed at various positions on a track parallel to the waveguide. By remotely activating or releasing them, pinching antennas can be activated at different locations as desired.} Since a single user is served in each time slot, the antenna can be positioned at the location closest to the $n$th user in the $n$th time slot, thereby minimizing propagation loss. 
Figs. \ref{fig_SWSA} (c) and (d) show the data rates achieved by conventional-antenna and pinching-antenna systems as functions of the user location in a rectangular area, respectively.  As can be seen from the figure, the use of a pinching antenna results in a significant performance gain over the conventional-antenna case, especially when the user is located far away from the center along the $y$ axis.

{ Second,  a pinching antenna may serve a group of users simultaneously, as shown in Fig. \ref{fig_SWSA} (b). The motivation of this scheme is that the number of  pinching antennas available for activation is smaller than the number of users, which is similar to the conventional MIMO situation, where the number of antennas is smaller than the number of users.}
With the second scheme, a pinching antenna is activated at a position directly above the user group, allowing it to serve multiple users simultaneously, where the challenge is to find a tunable antenna position to balance the trade-off between system throughput and user fairness. We note that this scheme can be particularly useful for multicast data transmission, i.e., users in the same group are to receive the same information. For the multicast case, OMA can be used to serve multiple groups in different time slots, while for the unicast case, hybrid OMA can be used, e.g., different time slots are assigned to different groups and different subcarriers are used to serve multiple users in one group.


\subsection{Single Waveguide with Multiple Activated Pinching Antennas}
Given the flexibility and low cost of pinching antennas, it is natural to consider the { simultaneous activation of multiple pinching antennas} on a single waveguide. In this case, the electromagnetic signals propagating along the waveguide will leak from the multiple antennas, yielding a ``single waveguide with multiple activated pinching antennas''  (SWMAP) architecture, as shown in Fig. \ref{fig_SWSA} (e). In SWMAP, there is a single feeder for the waveguide, and thus a single stream of data is passed through the waveguide. Note that there is a phase shift between the radiated signal from two adjacent pinching antennas caused by the waveguide propagation. In particular, the difference between the phases of the signals radiated by the $n$-th pinching antenna and the first pinching antenna is given by $\theta_n=2\pi(x_n-x_1)/\lambda_g$, where $x_n$ is the coordinate position of the $n$-th user. 
The above feature of pinching antenna systems is similar to conventional hybrid beamforming systems with a single radio frequency (RF) chain. However, unlike conventional hybrid beamforming, the positions of the pinching antennas can be adjusted so that the beamforming coefficients and channel parameters, such as free-space phase shifts and large-scale path loss, can be dynamically tuned, making the effective channel gains of the users reconfigurable. 

\begin{figure*}[!t]
\centering
{\includegraphics[width=0.9\linewidth]{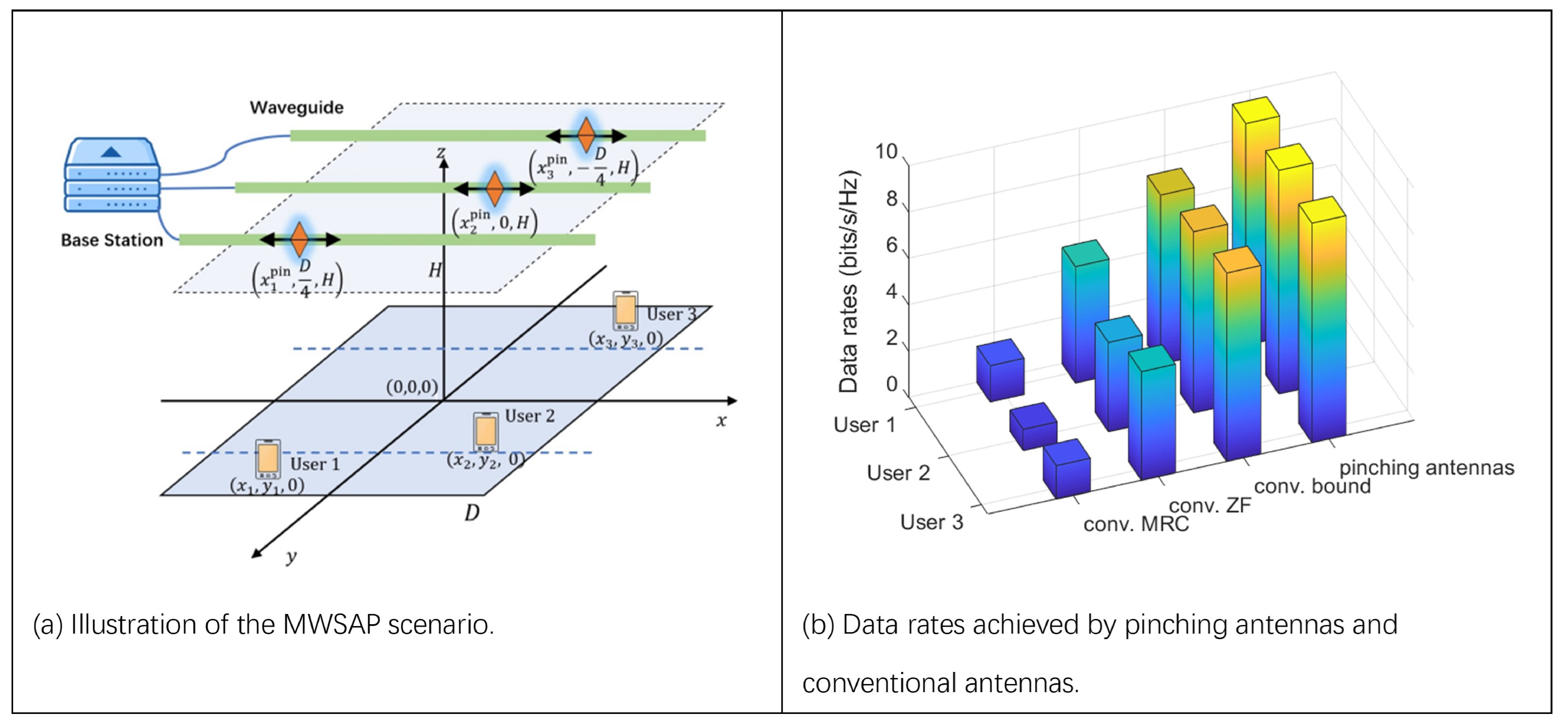}}
\label{fig_3userScenario}
\vspace*{-0.0em}
\caption{(a) Illustration of the MWSAP scenario, and (b) data rates achieved by multiple pinching-antenna systems and conventional-antenna systems.}
\label{compare_conv}
\end{figure*}

\section{Communications with Multiple Waveguides}

To further exploit the potential of pinching antennas, it is important to use multiple waveguides, where each waveguide can be equipped with one or more pinching antennas, resulting in the so-called multiple waveguides with a single {activated} pinching antenna (MWSAP) and multiple waveguides with multiple {activated} pinching antennas (MWMAP) architectures.
Note that each waveguide is fed by a single RF chain. Therefore,  digital beamforming can be applied in MWSAP/MWMAP systems, effectively exploiting the degrees of freedom (DoF) in the spatial domain. In particular, pinching antennas can be exploited  to implement MIMO and non-orthogonal multiple access (NOMA) in an innovative manner, as discussed in the following subsections. 

\subsection{New Forms of MIMO with Reconfigurable Channels}

The use of pinching antennas leads to a new type of MIMO system where not only the transceivers but also the MIMO channels can be reconfigured, as shown in the following example. Recall that in conventional MIMO systems, the performance of digital beamforming strategies is highly dependent on the intrinsic structure of the space formed by the user channel vectors. For example, zero-forcing (ZF) and maximum-ratio combing (MRC) are two commonly used principles for multi-user MIMO systems \cite{tse2005fundamentals}. 
The key idea of ZF is to set the beam vector for a given user in the null space of the channels of the other users with whom it would interfere, 
so that interference can be avoided, while the key idea of MRC is to set the beam vector for a given user in the same direction as the user's channel to maximize the strength of the intended signal. The holy grail of MIMO design is to realize a performance ceiling where the goals of ZF and MRC are simultaneously achieved. The reason for this challenge is the intrinsic dilemma between ZF and MRC: Minimizing the interference may also weaken the strength of the intended signal, while maximizing the strength of the intended signal may cause stronger interference. 

Interestingly, by using multiple pinching antennas on multiple waveguides, the channel gains  can be adjusted so that the structure of the users' channel space is also adjusted, which can significantly increase the possibility of achieving the aforementioned ideal upper bound. {As an example, consider a scenario as shown in Fig. \ref{compare_conv}(a), where three users are randomly distributed in a square, and three waveguides each equipped with a single pinching antenna are parallel to each other.} 
The results for the ``pinching antennas'' shown in Fig. \ref{compare_conv} (b) are obtained by optimizing
the positions of the pinching antennas on each waveguide. 
For comparison, a conventional MIMO antenna system is also considered where the three antennas are placed at $(-\frac{\lambda_0}{2},0,H)$,
$(0,0,H)$ and $(\frac{\lambda_0}{2},0,H)$, respectively, and $H$ is the height of the waveguide relative to the $x-y$ plane. Note that the results for the ``conv. bound'' of the three users were obtained {with} $R_i^B=\log_2(1+\rho|\mathbf{h}_i|^2)$, 
where $\mathbf{h}_i$ is the channel vector of user $i$ for the conventional-antenna system, and $\rho$ is the transmit power. 
As can be seen in Fig. \ref{compare_conv} (b), the data rates achieved by ZF and MRC beamforming are much lower than the upper bounds of the conventional-antenna system. 
On the other hand, the data rates achieved {with}  pinching antennas are much higher than those of the conventional-antenna system, particularly even higher than the upper limit of the conventional-antenna system.
The above observations indicate the great potential of pinching antennas for future wireless communication networks. 

\subsection{NOMA-Assisted Pinching-Antenna Systems}

As mentioned above,  pinching-antenna systems can outperform conventional MIMO systems through joint beamforming and location optimization. However, {their} performance is inherently limited by the number of waveguides and the number of pinching antennas on each waveguide.
In this context, NOMA can be applied to further enhance the performance of pinching-antenna systems, especially in the case where the number of waveguides is smaller than the number of users. The  users can be grouped into different clusters. In each cluster, the users can share the same beam formed by the digital beamformers and the locations of the pinching antennas, {while their signals can be separated by exploiting the power domain}

In conventional antenna systems, NOMA is usually implemented by ordering users according to their channel gains, where power allocation is also incorporated to improve user fairness by allocating more power to users with weak channel conditions. However, the capacity region and user fairness are fundamentally limited by the users' channel gains, which are not adjustable in conventional-antenna systems. Promisingly, in pinching-antenna systems, the effective channel gains of the users can be adjusted, providing more flexibility for performance improvement. For example, the order of the channel gains in pinching-antenna systems can be intelligently adjusted, so that more personalized service can be provided to improve user fairness.

\begin{figure*}[htbp]
\begin{center}
\includegraphics[width=0.9\linewidth]{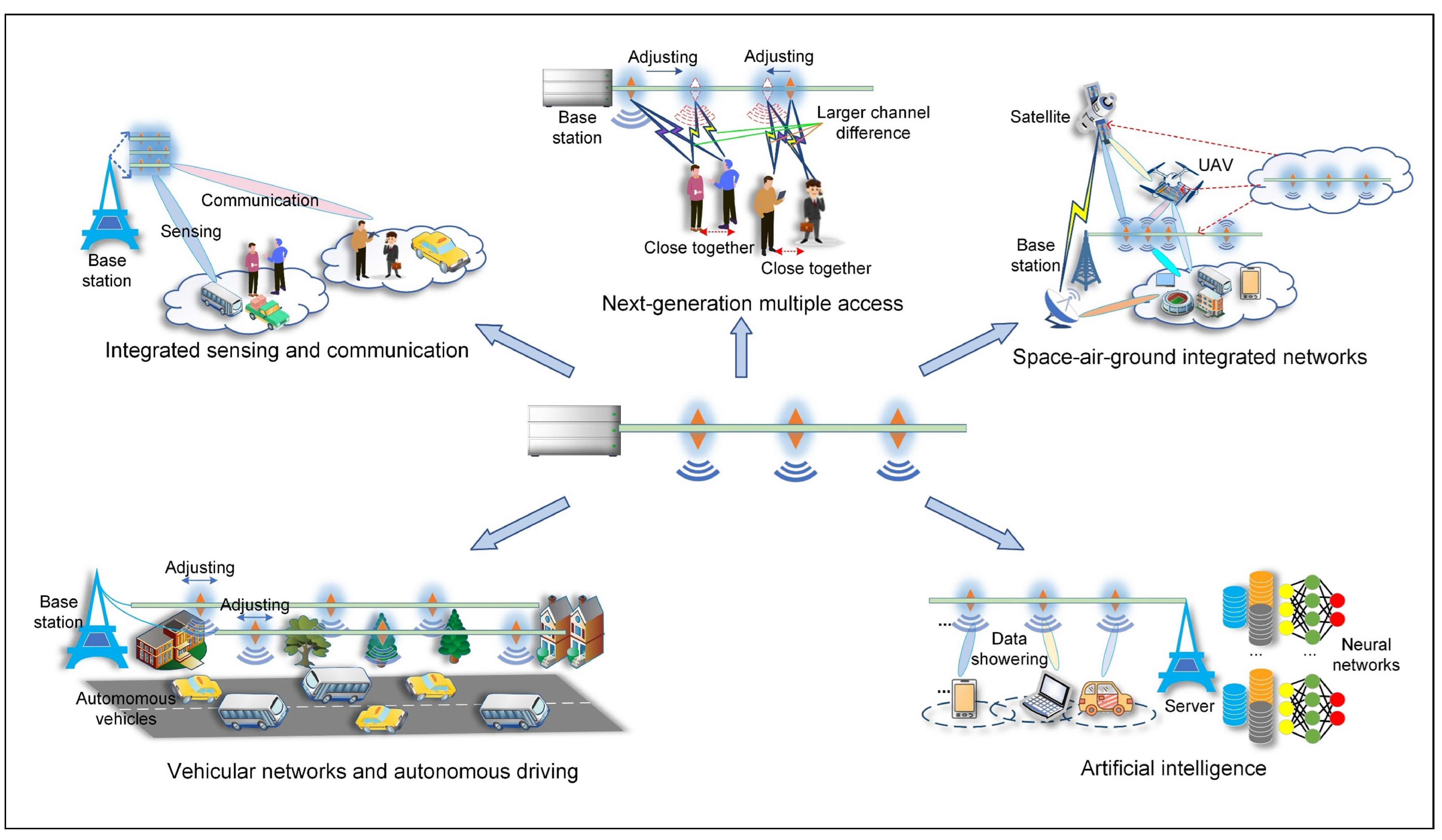}
\end{center}
 \vspace*{-0em}
\caption{ Promising applications of pinching-antenna systems.}
\label{fig_appsce}
\end{figure*}

\section{Promising Applications of Pinching-Antenna Systems}

Important 6G-related applications of pinching-antenna systems are shown in Fig. \ref{fig_appsce}, and are  discussed in the following subsections.

\subsection{Pinching-Antenna Assisted ISAC}	

Integrated sensing and communication (ISAC) is expected to play a critical role in IMT-2030 \cite{Prelcic24ISAC6G}, effectively using the LoS path to integrate sensing and communication functions.  As discussed in Section II, the use of pinching antennas can create new LoS links or enhance existing ones, which can be particularly beneficial for  the implementation of ISAC.  In addition,  {by increasing
the distance between the pinching antennas (i.e., anchors) and establishing strong LoS
links, the dynamic positioning and angular adjustment capabilities of pinching antennas can overcome the disadvantage of positioning methods  based on
conventionally packed antennas.}
This is especially beneficial in complex environments where multi-path effects or reflections can  distort or obscure the desired signal, making it difficult to maintain reliable communication. 


\subsection{Pinching-Antenna Assisted NGMA}	

The goal of next-generation multiple access (NGMA) is to overcome the limitations of traditional multiple access technologies in high-density, large-scale, low-latency environments through the introduction of innovative multiple access methods, ultimately improving system capacity, and spectrum efficiency \cite{NGMA24DING}. However, taking NOMA as an example, the implementation of NOMA can be challenging when users are in close proximity.  By flexibly adjusting the deployment of  waveguides and dynamically modifying the locations of  pinching antennas, distinct spatial beams can be created to differentiate users who are in close proximity of each other. In addition,  by flexibly adjusting the placement and phase shifts of pinching antennas, sophisticated control over signal strength can be achieved, ensuring efficient power allocation for users with weaker channels.

\subsection{Pinching-Antenna Assisted SAGIN}

%
Space-air-ground integrated network (SAGIN) have been considered as a promising architecture for 6G communication networks \cite{Guorongiotj24sagin}, primarily relying on satellite platforms, airborne platforms, and ground stations. These networks rely on LoS transmission for reliable, high-quality global coverage and efficient connectivity.   However, in complex terrains such as mountainous or urban areas, LoS links can be blocked by mountains or buildings, resulting in signal attenuation and degraded communication quality. {Fortunately, the use of pinching antennas, mounted on waveguides installed along the edges of tall buildings, can improve SAGIN performance by creating strong LoS links with satellites, which are crucial to satellites communication networks.}
 Furthermore, pinching antennas can be mounted on unmanned aerial vehicles, allowing dynamic activation and repositioning of the antennas to establish LoS links that can quickly adapt to changing environments and requirements.


\subsection{Pinching-Antenna Assisted Vehicular Networks}

6G-enabled vehicular networks will serve as the foundation for intelligent transportation systems, driving progress towards safer, more efficient and environmentally sustainable mobility solutions.  
The main challenges in vehicular communications are caused by dynamic and static obstacles, such as other vehicles, buildings, and trees that can block LoS communication paths. {Given the fact that vehicles can move inside the infrastructures, e.g., roads and bridges, deploying waveguides along these infrastructures makes the applications of pinching antennas to vehicular networks a feasible and attractive communication solution. In particular, the dynamic activation of radiation points by pinching antennas facilitates stable and high-quality LoS communication links \cite{Kaidi24pinchingact}.}
In addition, for autonomous driving, the flexible beamforming and fast activation of pinching antennas ensure real-time connections between vehicles and infrastructure, supporting cooperative driving and precise route planning to increase safety and efficiency.

\subsection{Pinching-Antenna Assisted AI}

Large-scale artificial intelligence (AI) model updates and data synchronization pose significant challenges for the development and deployment of AI systems, particularly in environments that require continuous learning, real-time data processing, and continuous model improvement \cite{Nguyenlsl23}.  By dynamically adjusting the radiation positions of pinching antennas, these antennas can efficiently respond to changing data demands by directing resources to data-intensive nodes and hotspot areas. As a result, pinching-antenna systems can effectively support the challenging low-latency, high-bandwidth data showering required by future AI applications. In addition, with pinching antennas,  the radiating points and paths can be dynamically adjusted, ensuring stable and continuous connectivity for devices. 
  

\begin{figure*}[htbp]
\begin{center}
\includegraphics[width=0.9\linewidth]{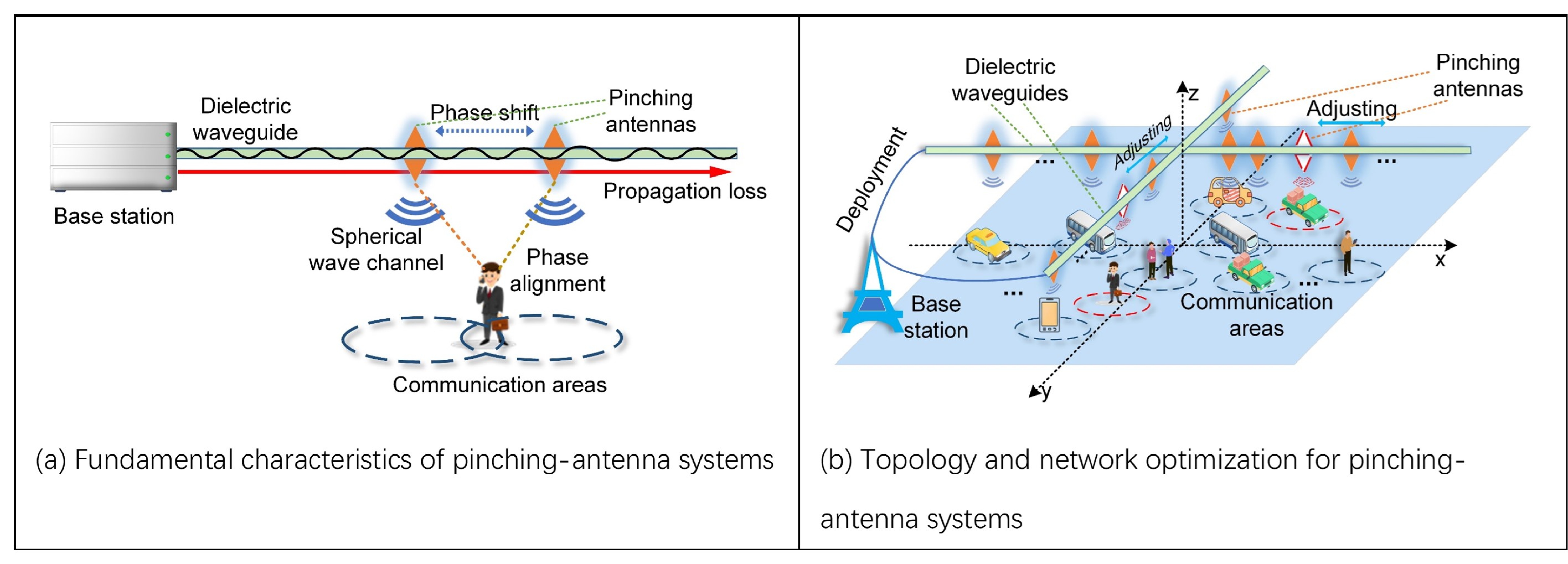}
\end{center}
 \vspace*{-0em}
\caption{Illustration of future directions for research.}
\label{Direfw}
\end{figure*}

\section{Directions for Future Research}

The design of pinching-antenna systems for wireless communications is  in its early stages of development, and thus, there are many important open issues and challenges, as discussed
 in the following subsections.

\subsection{Fundamental Characteristics of Pinching-Antenna Systems}

When multiple pinching antennas are activated  along a waveguide, propagation loss causes amplitude and phase variations, resulting in differences in the output signals from each antenna. This highlights the importance of developing a robust propagation loss model to accurately characterize the system's  behavior and effectively optimize its performance. In addition, there are two types of phase shifts that a user's signal experiences in pinching-antenna systems. One is caused by the signal propagating along the waveguide, and the other one is caused by the signal traveling to the user in free space. Phase alignment, i.e., {the compensation of these two phase shifts to ensure that the signals are combined constructively at the users' locations, is critical in pinching-antenna systems because accurate antenna placement is necessary to  match the phase shifts of the users' channels involve to achieve optimal performance. However, alignment errors can occur in practical environments}. Therefore, the joint optimization of antenna positioning, beamforming, and transmit power in the presence of potential phase alignment errors is an important direction for future research, as shown in Fig. \ref{Direfw} (a).

\subsection{Topology and Network Optimization for Pinching-Antenna Systems}


When users are evenly distributed over a circular or rectangular area, the use of multiple pinching antennas along with multiple waveguides is an effective strategy for improving network coverage. This approach requires optimizing both the number and placement of the waveguides while taking advantage of the dynamic adjustment capabilities of  pinching antennas to improve overall coverage and network performance. In particular, multiple waveguides can be placed in parallel or cross-pattern arrangements to enable centralized management and provide comprehensive coverage of  hotspot areas. Another approach is to deploy multiple waveguides at different locations in a distributed layout, which extends coverage over a larger area and enables more flexible network performance optimization. As a result, future research should focus on joint optimization of  the placement of pinching antennas and waveguides to ensure system effectiveness, as shown in Fig. \ref{Direfw} (b).

\subsection{Channel Estimation in Pinching-Antenna Systems}

The acquisition of channel state information (CSI) is important for  transceiver design for pinching-antenna systems. An effective approach to obtain CSI between each pair of transmit and receive 
antenna is to employ  predefined pilots. However, channel estimation for pinching antennas poses new challenges: Each waveguide is fed by only a single RF chain, while it may carry  multiple pinching antennas. As a result, it is necessary to recover high-dimensional information from low-dimensional observations, which requires more research.  Another way to obtain CSI is beam training, which focuses on estimating the channel gains of predefined beam patterns.  However, beam training for pinching antennas faces new challenges: The pinching antennas may not be equally-spaced and their activation locations may be dynamically adjusted, which will significantly increase the beam training overhead. Therefore, it is important to investigate new codebook designs as well as corresponding efficient beam training methods for pinching-antenna systems in the future.

\subsection{Uplink Transmission Configuration of Pinching-Antenna Systems}

Current work on pinching antennas has mainly focused on supporting downlink transmission \cite{Zhiguo24pinchinga}, \cite{Xu24pinchinga}, while the development of uplink transmission is at an infancy stage \cite{Sotiris24pinchingup}. In uplink transmission, the phase shifts in the received signals at the base station, caused by both waveguide and free-space propagation, can cause inconsistencies in signal synthesis, especially in multi-user and multi-pinching antenna scenarios, where variations in phase shifts can significantly affect system performance. The accumulation of these phase shifts can cause misalignment between signals from different users, reducing the efficiency of signal power combining and hindering the base station's ability to effectively demodulate uplink signals. Therefore, it is essential to design accurate phase compensation mechanisms that can dynamically adjust the  phase shifts introduced by waveguide and free-space propagation and ensure intelligent alignment of multiple users' signals at the base station.


\subsection{Machine Learning (ML) for Pinching-Antenna Systems}

 ML can play an important role in enabling the application of pinching-antenna systems. For example, due to the existence of phase shifts (including free-space and waveguide propagation phase shifts) that are determined by the locations of the pinching antennas, the optimization problems for the placement of pinching antennas are typically non-convex, and thus, difficult to solve using conventional methods.
  For this purpose, ML can be applied to find a sub-optimal solution in a computationally efficient  way. 
  In addition,
 considering the dynamic changes of the communication scenarios, the locations of the pinching antennas need to be adjusted intelligently to ensure robust performance. One option for optimizing the trajectories of the pinching antennas is to use reinforcement learning. 


\subsection{Combing Pinching Antennas with Other Flexible Antennas}

In the high-density urban environments of smart cities, the distribution of users is highly dynamic and the presence of significant multi-path effects poses significant challenges to traditional communication systems. To address these challenges, the combination of pinching antennas with other flexible antenna systems, such as fluid and movable antennas, has significant potential. Distributed
pinching antennas connected to the base station can leverage their strong LoS capabilities to provide stable communications links in densely populated areas.
Users equipped with fluid or movable antennas can dynamically adjust both the location and direction of their antennas, thereby fully exploiting the spatial diversity gains. Therefore, it is worthwhile to study the synergy effects between  multiple flexible antenna systems, as this not only improves the quality of signal coverage, but also the overall network performance in high-density, complex urban environments.



\section{Conclusion}

In this article, we have first introduced the basic concepts of pinching antennas and illustrated their key features of providing strong LoS links and enabling flexible antenna array configurations. Then, the implementation of pinching antennas has been explored  using a single waveguide, and adopting configurations with single and multiple activated pinching antennas, respectively. Furthermore,  the advantages of joint beamforming design and position optimization have been investigated for  pinching-antenna systems employing multiple waveguides, and the benefits  of applying NOMA in pinching-antenna systems have been unveiled. In addition,  the potential integration of pinching antennas into various emerging 6G application scenarios  have been discussed, including ISAC, NGMA, SAGIN, vehicular networks, and AI. Finally, directions for future research have been highlighted.



\bibliographystyle{IEEEtran}
\bibliography{IEEEabrv,ref}
\end{document}